\documentclass[journal]{IEEEtran}

\usepackage{amsthm}
\usepackage{multirow}
\usepackage{style/commands}
\usepackage{style/vk}
\usepackage{booktabs}
\usepackage[normalem]{ulem}

\ifCLASSINFOpdf
\else
\fi
\begin{document}

%
\title{Incidents Are Meant for Learning, Not Repeating: Sharing
  Knowledge About Security Incidents in Cyber-Physical Systems}
%
%
%

\author{Faeq Alrimawi, Liliana Pasquale, Deepak Mehta, Nobukazu
  Yoshioka, Bashar Nuseibeh
  
  \thanks{Place acknowledgements here.}}

\maketitle

\begin{abstract} Cyber-physical systems (CPSs) are part of most
  critical infrastructures such as industrial automation and
  transportation systems. Thus, security incidents targeting CPSs can
  have disruptive consequences to assets and people.  As
  prior incidents tend to re-occur, sharing knowledge about these
  incidents can help organizations be more prepared to prevent, mitigate or
  investigate future incidents. 
  This paper proposes a novel approach to enable representation and sharing
  of knowledge about CPS incidents across different organizations. To
  support sharing, we represent incident knowledge (\emph{incident
    patterns}) capturing incident characteristics that can manifest
  again, such as incident activities or vulnerabilities exploited by
  offenders. Incident patterns are a more abstract representation of specific 
  incident instances and, thus,  are general enough to be applicable to
  various systems - different than the one in which the incident
  occurred. They can also avoid disclosing potentially
  sensitive information about an organization's assets and
  resources. We provide an automated technique to \emph{extract} an incident
  pattern from a specific incident instance. To understand how an incident
  pattern can manifest again in other cyber-physical systems, we also provide an automated technique
  to \emph{instantiate} incident patterns to specific systems. We demonstrate the feasibility of our
  approach in the application domain of smart buildings. We evaluate
  correctness, scalability, and performance using two substantive
  scenarios inspired by real-world systems and incidents.
\end{abstract}


\section{Introduction}
\label{sec:introduction}

Cyber-Physical Systems (CPSs) combine computation, communication, and
physical processes~\cite{Lee.2010} to augment physical systems with enhanced
capabilities, such as real-time monitoring and dynamic control.
Nowadays applications of CPS can be found in a multitude of domains
such as industrial control systems, transportation systems and smart
buildings. Thus, security incidents targeting CPSs can have disruptive
consequences to their users and the assets they manage.

CPSs enable complex interactions between cyber and physical
components. For example, in a smart building, a rise in the measured
temperature of a room can trigger a digital process to issue a command
to an air conditioner to start cooling the room. Interactions between
cyber and physical components can extend the attack surface of a CPS,
giving malicious individuals more opportunities to cause harm since
they can exploit vulnerabilities from either component to impact the
other. Thus, the number of security incidents targeting CPSs
has increased over the past years~\cite{Loukas.2015}. For example, in the
Ukrainian power grid incident~\cite{UkrainianAttack}, offenders used spear phishing
to gain a foothold in the distribution companies' computer network.
Then, they gained access to the power grid network, to infect and
disable some physical devices (e.g., workstations, serial-to-Ethernet)
that control the electricity distribution, causing disruption to the
normal operation of the grid. Previously, in the German steel-mill
incident~\cite{GermanSteelMillAttack}, offenders used spear phishing to gain a foothold in
the corporate network, and then gain access to the plant’s network in order to shutdown the blast furnace
and the alarm system.

Some aspects of prior incidents often tend to re-occur. For example, in
the Ukrainian power grid incident an offender obtained access to a
private network using spear phishing. A similar activity also
happened in the German steel-mill incident. Thus, sharing knowledge about prior
security incidents can help organizations being more prepared to
prevent, mitigate, or investigate future
incidents~\cite{Cardenas.2009}. However, supporting information
sharing about security incidents is still a key open challenge~\cite{NIST800-61,Schreck.2018}. Besides, knowledge about security incidents
targeting CPSs is limited.

In this paper we propose a novel approach to enable representation and
sharing of incident knowledge across different organizations. To
enable sharing, we represent incident knowledge (\emph{incident
  patterns}) capturing incident characteristics that can manifest
again, such as incident activities or vulnerabilities exploited by
offenders. Incident patterns are a more
abstract representation of  specific incident instances. Thus they are
general enough to be applicable to
various systems - different than the one in which the incident
occurred. They can also  avoid  disclosing potentially sensitive information
about an organization's assets and resources
(e.g., physical structure of a building or vulnerable
devices).  As incident activities can target or exploit system components, we also
provide a representation of the system where an incident occurs.
This includes  cyber and physical components, their structure, 
dependencies and dynamic
behavior.  We provide two meta-models to represent incidents and cyber-physical systems,
respectively.

We propose an automated technique to \emph{extract} an incident pattern from a specific
incident instance.  The extraction technique explores inheritance
hierarchy in the cyber-physical system meta-model to abstract specific
system characteristics described in the incident instance. To
understand how an incident pattern can manifest again in other CPSs,
we propose an automated technique to \emph{instantiate} incident patterns to different
systems. The instantiation technique uses a representation of the
dynamic behavior of the system - expressed as a labeled transition
system - to identify the behavior traces matching the activities in
the incident pattern. We demonstrate the feasibility of our approach
in the application domain of smart buildings. We evaluate correctness,
scalability, and performance using two substantive
scenarios inspired by real-world systems and incidents.

The novelty of our work lies in the combination of three key elements:
\begin{itemize}
\item Incident patterns to support representation and sharing of incident knowledge across different organizations.
\item Automated technique to extract an incident pattern from
  a specific incident instance in order to facilitate sharing of incident
  knowledge and avoid disclosing sensitive information about an
  organization's assets and resources.
\item Automated technique to instantiate an incident
  pattern  to specific cyber-physical systems in order to facilitate
  assessment about whether and how prior incidents can manifest again.
\end{itemize}

The remainder of this paper is organized as follows. In
Section~\ref{sec:motivating-example}, we motivate the need to share
information about incidents in CPSs. In
Section~\ref{sec:shar-incid-knowl}, we provide an overview of our
approach to share incident knowledge. In section~\ref{sec:model-sys-inc}, we
present, respectively, our system and incident meta-models. In Section~\ref{sec:extraction}, we describe the incident pattern extraction technique. In Section~\ref{sec:instantiation}, we illustrate the incident pattern instantiation technique. In
Section~\ref{sec:evaluation}, we present an evaluation of both techniques, and discuss
the results and threats to validity. In section~\ref{sec:relatedWork},
we compare our approach with
related work. Finally, in Section~\ref{sec:conclusion}, we conclude and present future
work.


\section{Motivating Example}
\label{sec:motivating-example}

\begin{figure}[htbp]
  \centering
  \includegraphics[width=0.7\columnwidth]{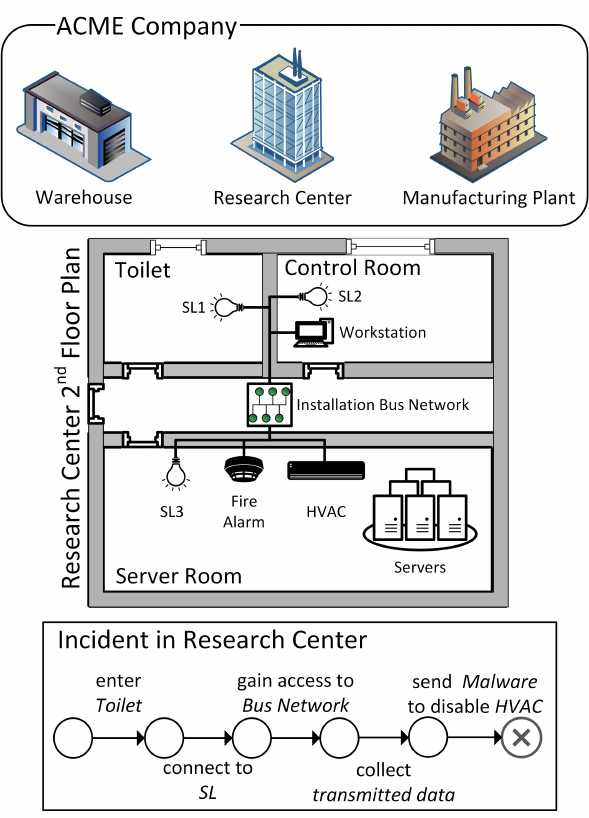}
\caption{The ACME Company Motivating Example.}
\label{fig:example}
\end{figure}

Our motivating example is centered on the ACME
company that operates across three different smart buildings: a
\emph{Research Center}, a \emph{Warehouse}, and a \emph{Manufacturing
  Plant}. This is depicted in Fig.~\ref{fig:example}. The plan of the 2nd floor of the Research Center consists of
a \emph{Server Room}, a \emph{Control Room}, and a \emph{Toilet}. The
\emph{Server Room} contains a \emph{Fire Alarm}, an air conditioning
unit (\emph{HVAC}), and some \emph{Server}s, while the \emph{Control
  Room} contains a \emph{Workstation}. The whole building is equipped
with\emph{ Smart Light}s. The \emph{HVAC}, the \emph{Fire Alarm} and
the \emph{Smart Light}s communicate with the \emph{Workstation}
through the\emph{ Installation Bus} network, which adopts the KNX
protocol~\cite{KNX}.

Unfortunately, an incident occurred in the \emph{Research Center}. An
offender reached the 2nd floor, entered the \emph{Toilet}, and
connected to the \emph{Smart Light} (\emph{SL1}) using a Laptop. After
that, s/he obtained access to the \emph{Installation Bus} and was able
to collect data transmitted over the network (e.g., data exchanged
between the \emph{Workstation} and the \emph{HVAC}). This was possible
because messages exchanged through the installation bus are not
encrypted~\cite{Mundt.2016}. The offender then sent a targeted Malware
(e.g., exploiting the vulnerabilities present in Trane
HVACs~\cite{Krebs2016}) to disable the \emph{HVAC}, subsequently
causing the \emph{Servers} to heat up. The incident actions are listed
at the bottom of Fig.~\ref{fig:example}.

Upon the discovery of the incident, security administrators wrote a
report describing how the incident occurred. Afterwards, to assess whether similar
incidents activities can manifest in the other buildings,
security administrators have to identify existing vulnerabilities brought by cyber and
physical components in those buildings. This 
may require to examine the physical structure of each building, as well as the software and
network configurations of the digital devices within the buildings. 

  However, security administrators face the following challenges. First, the approach each organization follows to perform incident reporting is
  not standardized~\cite{Ahmad.CS.2012}. Although different templates
  (e.g.,~\cite{Handbook}) have been proposed to identify what type of
  information incident reports should contain, the description of an
  incident is usually provided in natural language~\cite{Tulechki.2015}. Understanding how
  an incident can re-occur in a different system is arduous because it
  would require a security administrator to examine the incident
  description manually and speculate on all possible ways in which
  incidents activities can be performed. Second, incident reports may
  not represent incident activities in CPSs, which can exploit
  vulnerabilities and dependencies between cyber and physical
  components. Finally, incident reports may contain sensitive
  information (e.g., internal network structure) that cannot be
  disclosed to third parties~\cite{Ahmad.CS.2012}.  

Therefore, it is necessary to provide an approach to represent
incidents in a  more structured form, which can capture activities that
can use cyber and physical components.
Moreover, it is necessary to provide a modeling technique that could
allow representing incident information in an abstract form which avoids
disclosing specific sensitive information about the system in which
the incident occurred. Finally, automated techniques should be
provided to analyze incident knowledge and assess whether and how
prior incidents can re-occur.


\section{Sharing Incident Knowledge}
\label{sec:shar-incid-knowl}

\begin{figure}[htpb]
\centering
\includegraphics[width=0.99\columnwidth]{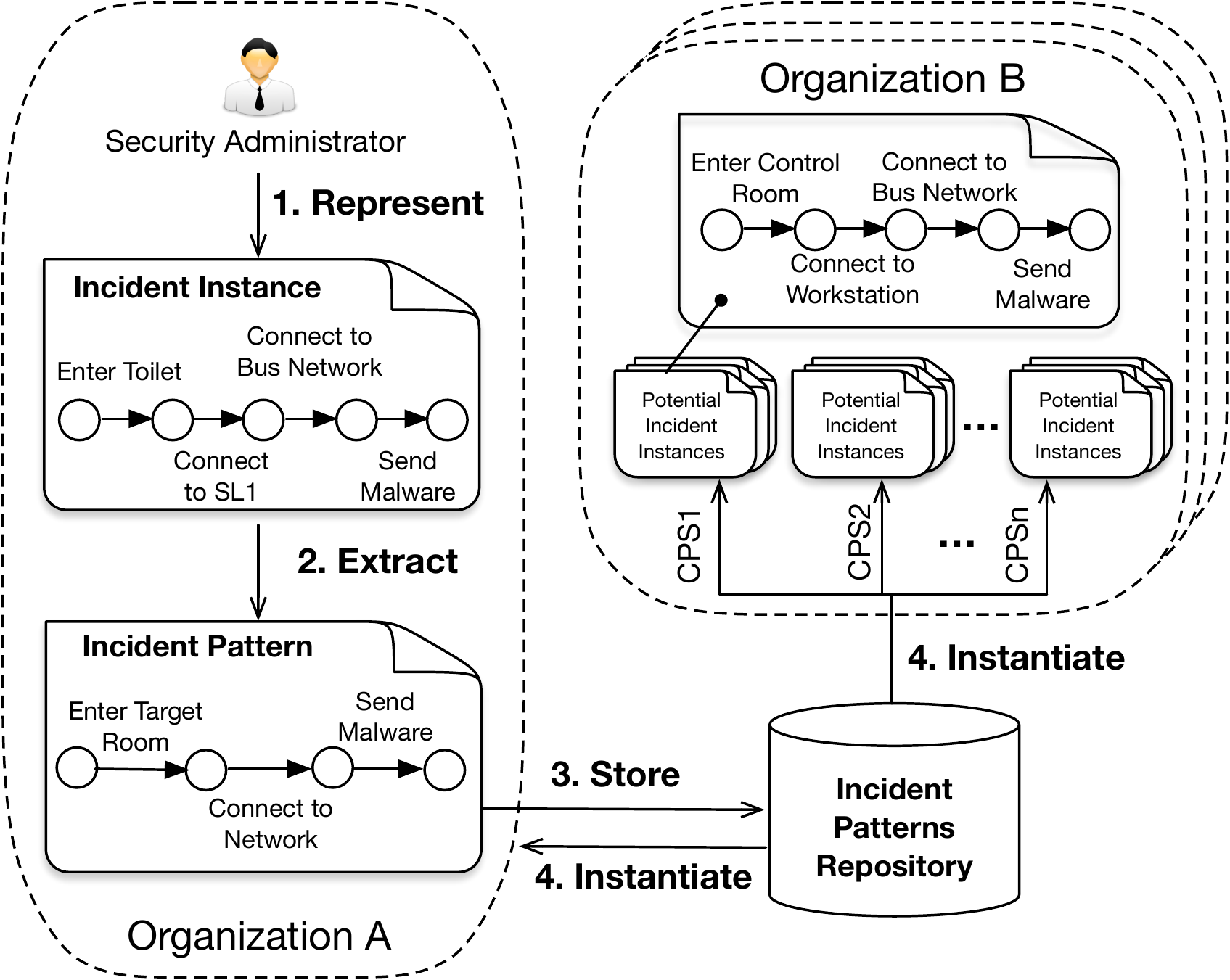}
\caption{Our approach for sharing incident knowledge}\label{fig:overall}
\end{figure}

To address the challenges highlighted in the previous section, we
propose our approach to share incident knowledge across different
systems and organizations. Our approach provides representations of
incidents and the cyber-physical systems in which they can occur. It
also provides two automated techniques. One technique  extracts
potential incident patterns from specific incidents. The other technique instantiates incident patterns in CPSs in order to assess whether and how an incident can reoccur. Our approach is shown in Fig.~\ref{fig:overall}.

After an incident occurs a security administrator operating within an
organization (\emph{A}) can \emph{represent} (1) the  activities of an incident (\emph{incident
  instance}). These activities refer to specific cyber and
physical system components (e.g., \emph{SL1} or \emph{Bus Network of
  the ACME Research Centre}). Subsequently, an \emph{incident pattern}
can be \emph{extracted} (2) from an incident instance automatically.
An incident pattern includes activities that refer to a more abstract
representation of cyber-physical system components. For example,
activity \emph{Enter Toilet} in the incident instance can be
abstracted to \emph{Enter Target Room} in the incident pattern. This
avoids disclosing specific information indicating, for example,
location of smart devices exploited in the incident. Differently from the traditional notion of
\emph{pattern}, in this paper an incident pattern is extracted from
a single incident instance. However, an incident pattern can 
potentially be instantiated in various ways in different systems.

The incident
pattern is subsequently \emph{stored} (3) in a shared repository. Each
time the repository is updated, a set of subscribed organizations is
notified. They can access incident patterns and \emph{instantiate} (4)
them automatically w.r.t a set of specific CPSs they manage. This
allows system administrators to assess all possible ways in which
incident patterns can re-occur again in those CPSs. For example, the
incident pattern extracted from the Research Center incident can be
instantiated to another organization's CPS (e.g., \emph{CPS1 managed
  by organization B}). Incident activity \emph{Connect to Network} in
the incident pattern is
instantiated to 2 subsequent activities \emph{Connect to Workstation}
and \emph{Connect to Bus Network}.

In our previous work~\cite{Alrimawi.SEAD.18}, we briefly introduced
our approach for sharing incident knowledge and provided a preliminary
description of the models adopted to represent an incident and the
cyber-physical system in which it can occur. In this paper, we extend our previous
work, by describing in more detail our models and how they can be used
to represent cyber-physical systems, incident instances and patterns. Differently
from our previous work, in this paper we  also provide two automated
techniques to support extraction and instantiation of incident
patterns, respectively. Finally we evaluate our techniques on a substantive, large-scale
example.


\section{MODELING SYSTEMS \& INCIDENTS}
\label{sec:model-sys-inc}

To model incidents in cyber-physical systems, we provide two meta-models. First, a cyber-physical system meta-model represents CPSs where  an incident can occur, focusing on their components, structure and dynamic behavior. Second, an incident meta-model is proposed to represent incident patterns and instances. 
The full meta-models are implemented as Eclipse plugins that are available publicly\footnote{https://tinyurl.com/yd9k6zhe}.

\subsection{Modeling Cyber-Physical Systems}
We tailor our system meta-model to represent smart buildings, which are a specific application domain of CPSs. So far the analysis of security incidents in smart buildings~\cite{Mundt.2016} has received little attention and this has motivated our focus on this domain.

\begin{figure}[htpb]
	\centering
	\includegraphics[width=0.98\columnwidth]{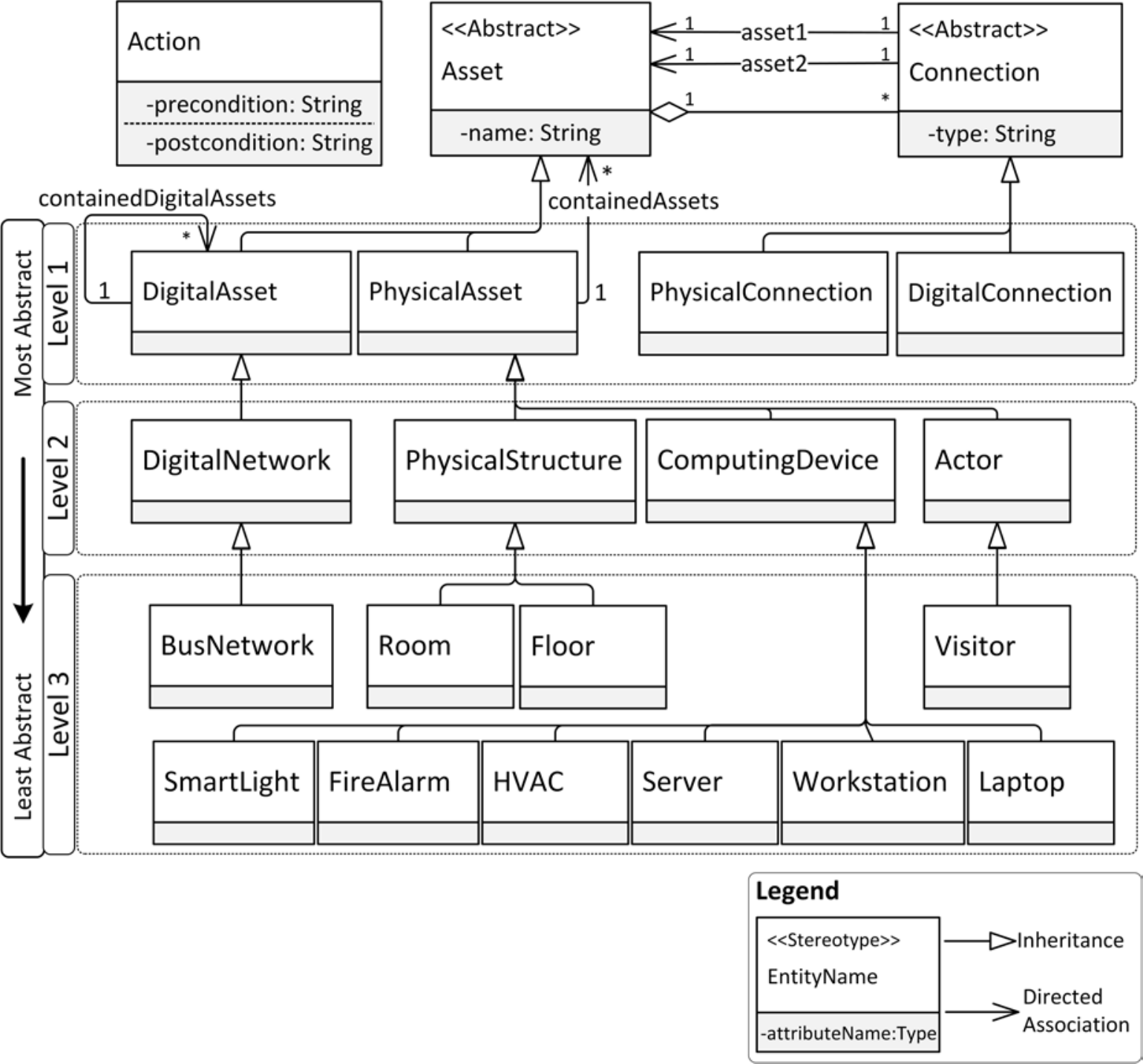}
	\caption{Smart Building meta-model (simplified).}\label{fig:sysMM}
\end{figure}

A simplified version of the smart building meta-model is shown in Fig.~\ref{fig:sysMM}. Note that this meta-model can be extended to represent other time-discrete CPSs in domains different than smart buildings. Our meta-model includes \emph{Asset}s, which can represent physical and cyber components in a smart building. 
Each Asset instance is identified by its \emph{name}. 
\emph{PhysicalAssets} represent any physical component, such as \emph{Actor}, \emph{PhysicalStructure}, and \emph{ComputingDevice}. \emph{Actor} can be a person in the smart building such as a \emph{Visitor} or an \emph{Employee}. 
\emph{PhysicalStructure} represents part of the smart building physical layout, which includes \emph{Room} and \emph{Floor}. \emph{ComputingDevice} represents any computing device, such as \emph{Laptop}, \emph{FireAlarm}, \emph{SmartLight}, \emph{Server}, \emph{HVAC}, and \emph{Workstation}. 
\emph{DigitalAsset}s represent any digital data that can be processed or software that can be installed in a digital device inside the smart building. A DigitalAsset can also represent a network (e.g., \emph{BusNetwork}) installed in the smart building.

\begin{figure}[htpb]
	\centering
	\includegraphics[width=0.90\columnwidth]{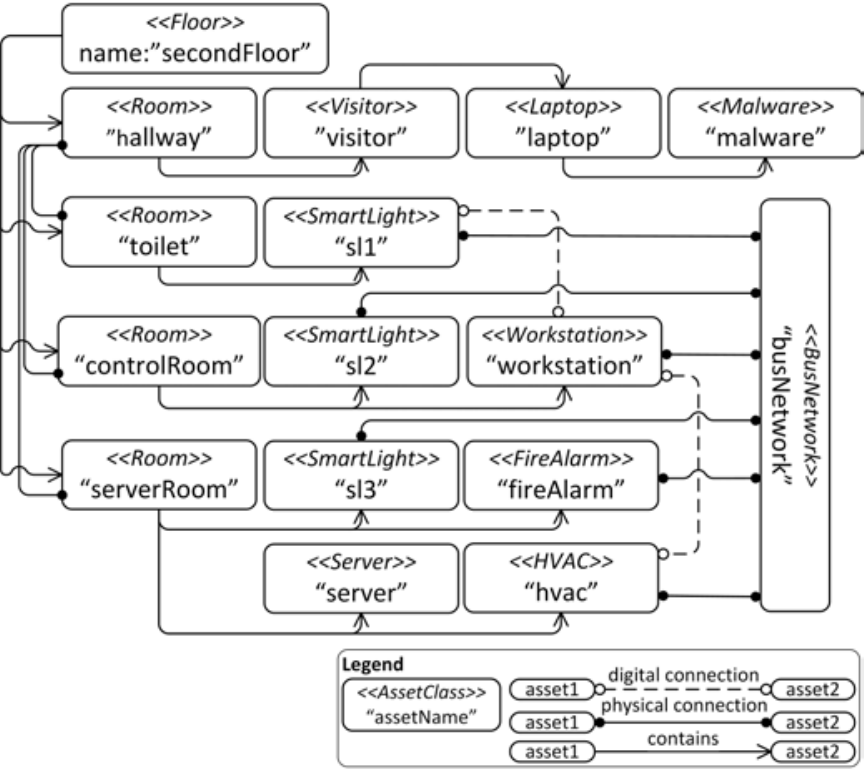}
	\caption{Research Center instance (simplified).}
	\label{fig:sysInstance}
\end{figure}

An instance of the smart building meta-model representing the Research Centre of the ACME Company is shown in Fig.~\ref{fig:sysInstance}. For example, \emph{sl1} and \emph{toilet} are instances of \emph{SmartLight} and \emph{Room}, respectively.

To model the structure of a CPS, the meta-model also represents containment and connectivity relations between components. 
The \emph{containedAssets} relation denotes the Asset(s) contained in a PhysicalAsset. The \emph{containedDigitalAssets} relation denotes the DigitalAsset(s) contained in another DigitalAsset. For example, as shown in Fig.~\ref{fig:sysInstance}, \emph{sl1} is contained in the \emph{toilet}, \emph{sl2} and the \emph{workstation} are contained in the \emph{controlRoom}, and \emph{sl3}, the \emph{fireAlarm}, the \emph{server} and the \emph{hvac} are contained in the \emph{serverRoom}. \emph{Connection} represents connectivity between two components (\emph{asset1} and \emph{asset2}) and can be described by a \emph{type} (e.g., wired). Digital connectivity between assets (e.g., through a network) is expressed as a \emph{DigitalConnection}, while physical connectivity between assets (e.g., two rooms are connected through a door) is expressed as a \emph{PhysicalConnection}.
For example, as shown in Fig.~\ref{fig:sysInstance},  the \emph{toilet} and the \emph{serverRoom} are connected physically to the \emph{hallway}, while \emph{sl1-sl3}, the \emph{fireAlarm} and the \emph{workstation} are connected physically to the \emph{busNetwork}. The \emph{workstation} is also connected digitally to the \emph{hvac} and the \emph{fireAlarm}, to which it sends control commands.

To specify the dynamic behaviour of a CPS, the meta-model allows representing \emph{Action}s.
For example, Actions can represent a person entering a room or connecting his/her laptop to a computing device via the bus network. An Action is expressed as a re-writing rule where a portion
of the system matching a \emph{pre-condition} is re-written with the sub-system represented in the \emph{post-condition}. 
Pre- and post-conditions are expressed using a custom notation inspired by Bigraphical Reactive Systems  (BRS)~\cite{Milner2006}, which allows representing cyber and physical components and their connectivity and containment relationships.

Table~\ref{tab:actions} represents pre-  and  post-conditions of actions ``\emph{enter Room}'', ``\emph{connect Laptop  to BusNetwork physically}'', and ``\emph{connect Laptop to ComputingDevice via BusNetwork}''.

\begin{table}[htbp]
	\footnotesize
	\caption{Pre- \& post-conditions of some actions of the smart building example.}
	\label{tab:actions}
	\begin{tabular}{ |p{0.2cm} l|  }
		\hline 
		\multicolumn{2}{|l|}{\textbf{enter Room}}\\\hline
		\textbf{pre:}& $(Room_1\{phys\}\cdot Actor)~|~(Room_2\{phys\})$ \\   
		\textbf{post:}& $Room_1\{phys\}~|~(Room_2\{phys\}\cdot Actor)$\\ 
		\hline
		\multicolumn{2}{|l|}{\textbf{connect Laptop (Lap) to BusNetwork (Bus) physically}} \\\hline
		\textbf{pre:}& $((Actor\cdot Lap)~|~Dev\{phys\})~||~Bus\{phys\}$  \\
		\textbf{post:}&$((Actor\cdot Lap\{phys\})~|~Dev)~||~Bus\{phys\}$ \\
		\hline
		\multicolumn{2}{|l|}{\textbf{connect Laptop to ComputingDevice (Dev) via BusNetwork}} \\\hline
		\textbf{pre:}& $Actor\cdot Lap\{phys\}~||~Bus\{phys\}~||~Dev\{phys,dig\} $  \\
		\textbf{post:}&$Actor\cdot Lap\{phys, dig\}~||~Bus\{phys\}~||~Dev\{phys, dig\}$ \\
		\hline  
	\end{tabular}
	
\end{table}


The precondition of Action ``\emph{enter Room}'' means that two different rooms ($Room_1$ and $Room_2$) are connected physically ($\{phys\}$) and are contained in the same physical structure (see operator '$|$'), for example, the same floor. An
$Actor$ is inside $Room_1$ (see operator '$.$'). As a result of Action ``\emph{enter Room}'', the $Actor$ who was
previously contained in $Room_1$ is now inside  $Room_2$.

Action ``\emph{connect Laptop to BusNetwork physically}'' indicates that an actor establishes a physical connection of a laptop to a bus network by replacing a computing device, which was previously connected to the bus network. As indicated in the pre-condition, an \emph{Actor}  who carries a laptop ($Lap$) (see operator '$.$') is initially co-located ('$|$') with a computing device ($Dev$). This device is in turn connected physically ($\{phys\}$)  to the bus network ($Bus$). Also $Actor$ and $Dev$ are not necessarily contained in the same location as $Bus$ (see operator '$||$'). In the post-condition $Lap$ is connected physically to $Bus$, replacing $Dev$.


Action ``\emph{connect Laptop to ComputingDevice via BusNetwork}'' indicates that an actor connects a laptop  digitally to a computing device through the bus network. The pre-condition indicates that an $Actor$ carries a laptop ($Lap$) (see operator '$.$'). The laptop ($Lap$) and the computing device ($Dev$) are connected physically ($\{phys\}$) to a bus network ($Bus$). The post-condition indicates that $Lap$ establishes a digital connection ($dig$) with $Dev$. Note that  $Actor$, $Bus$ and $Dev$ are not necessarily contained in the same physical structure  (see operator '$||$').


CPS components can be defined at different levels of abstraction in the meta-model. Level 3 includes concrete entities, while Levels 1 and 2 include more abstract entities. A CPS, such as the one represented in Fig.~\ref{fig:sysInstance}, is described by instances of the most concrete entities of the smart building meta-model (i.e. those in Level 3 in Fig.~\ref{fig:sysMM}). 



\subsection{Modeling Incidents}

\begin{figure}[htpb]
	\centering
	\includegraphics[width=0.99\columnwidth]{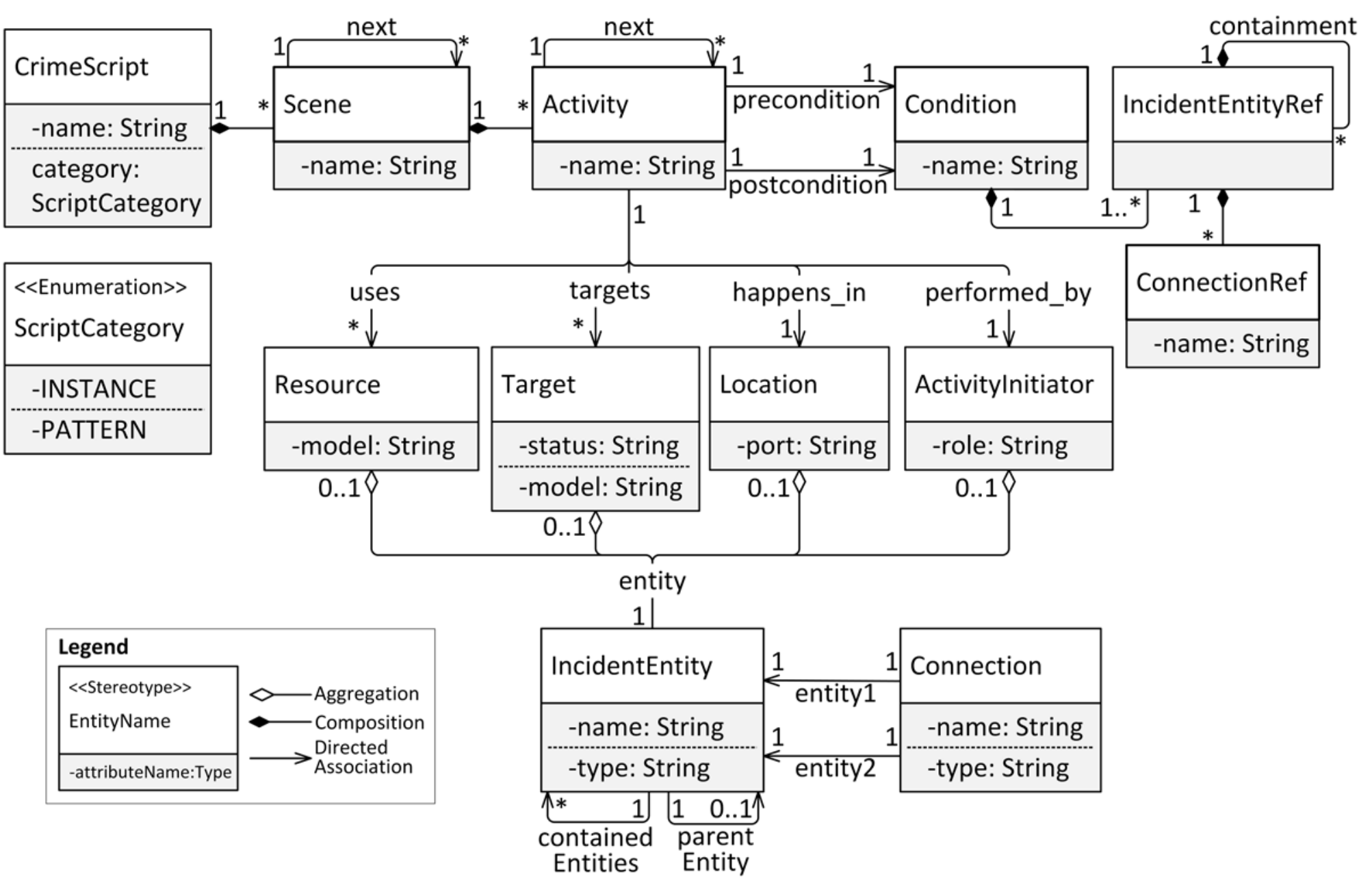}
	\caption{Incident meta-model (simplified).}\label{fig:incidentMM}
\end{figure}

We take inspiration from \emph{Crime Script}s to model security incidents. \emph{Crime Script}s are used in criminology to describe the sequence of activities of physical crimes~\cite{Cornish.1994}. Despite their adoption for
understanding the incident commission process and identifying incident prevention techniques, there exists no model that can be used to represent and process a \emph{Crime Script} systematically. Thus, we have developed a meta-model that
captures the characteristics of \emph{Crime Script}s. To represent incidents that occur in CPSs our meta-model extends the original use of \emph{Crime Script}s to refer to cyber components of the system explicitly. Our meta-model can be used to represent 
incident instances and incident patterns. An incident instance represents an incident that has occurred or may occur in a specific CPS, such as the \emph{Research Center} in our motivating example. Therefore incident instances can only refer to concrete CPS entities. An
incident pattern is a more abstract representation of an incident,
which can occur in various CPSs sharing common characteristics. Thus,
incident patterns  can  only refer to entity types (classes) of the CPS meta-model. 

A simplified version of the incident meta-model is shown in Fig.~\ref{fig:incidentMM}. A \emph{CrimeScript} entity is characterized by a \emph{name} and a \emph{category}. A category indicates whether the incident model represents an
incident \emph{INSTANCE} or a \emph{PATTERN}. A \emph{CrimeScript} includes a set of partially ordered \emph{Scenes}, which represent the phases of a security incident (e.g., preparation and execution scenes). Each scene, in turn, includes a set
of \emph{Activities}. An \emph{Activity} is characterized by a
\emph{name} and is linked to the next subsequent activities in chronological order. 

\begin{figure*}[htpb]
	\centering
	\includegraphics[width=\textwidth]{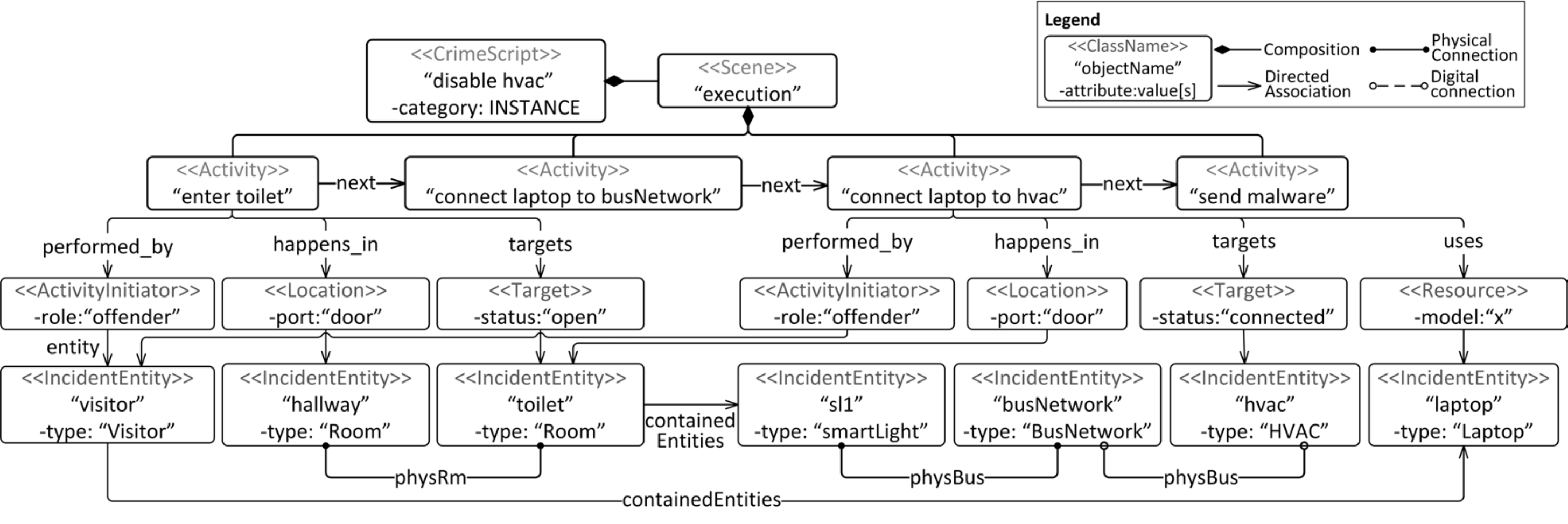}
	\caption{Incident instance model of the Research Center incident.}\label{fig:incidentInstance}
\end{figure*}

\begin{figure*}[htpb]
	\centering
	\includegraphics[width=\textwidth]{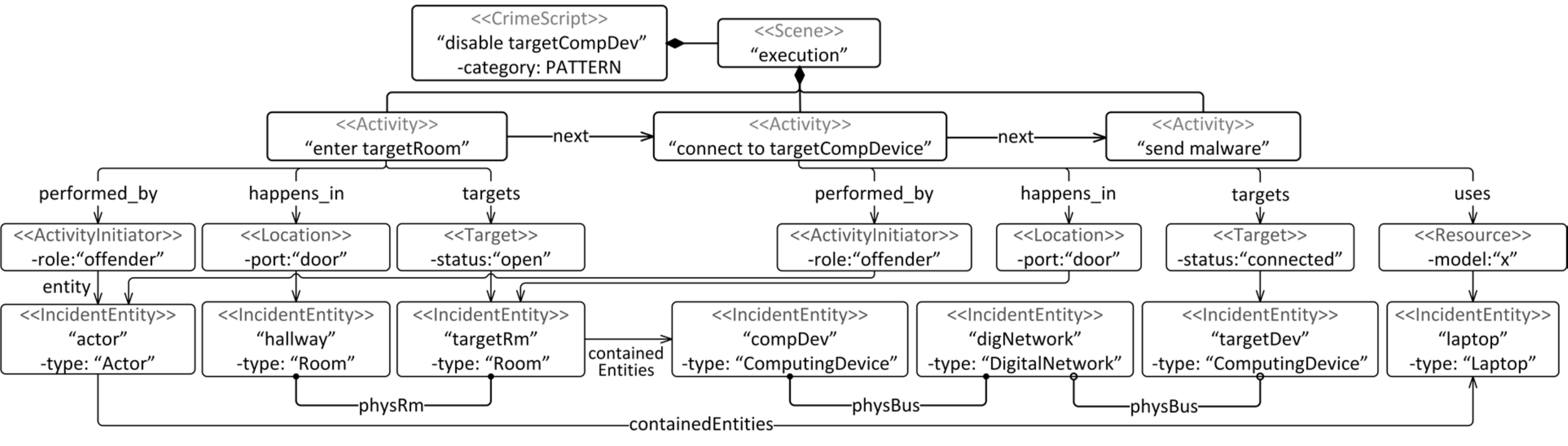}
	\caption{A potential incident pattern model extracted from the incident instance model.}\label{fig:incidentPattern}
\end{figure*}

An \emph{IncidentEntity} represents any entity that can be relevant to
an incident, such as an offender or an asset, and it is characterized
by a unique \emph{name} and  a \emph{type}. An \emph{IncidentEntity}
can ---not necessarily--- play different roles in an \emph{Activity}. It can be a
\emph{Resource}, a \emph{Target}, an \emph{ActivityInitiator}, or a
\emph{Location}. A \emph{Target} represents a component that can be
harmed during an incident. It is characterized by a \emph{status} (e.g., open, connected),
and a \emph{model} (e.g., Windows 10). A
\emph{Resource} represents a component needed to perform an activity,
such as a laptop or a malware, and is also characterized by a
\emph{model}. An \emph{Activity} can
be performed by an \emph{ActivityInitiator} who may have a
\emph{role} (e.g., offender, victim). A \emph{Location} represents a
place where an activity or a sequence of activities of a scene is
performed. A \emph{Location} can be physical or digital, depending on
the \emph{IncidentEntity} it refers to.  A physical location
represents a place in the physical space (e.g., a \emph{Room}). A digital
location represents a place in the cyberspace, such as an IP address
or a digital folder. A \emph{Location}'s \emph{port} defines an access
point to the location (e.g., a door for a physical location or port 80
for a digital location). An \emph{IncidentEntity} can contain other entities  through relation \emph{containedEntities}; the containing entity ---if present--- is represented through relation \emph{parentEntity}. A \emph{Connection} represents a physical or digital connectivity relation between two \emph{IncidentEntity} objects.  

An incident activity is also characterized by a  \emph{pre-} and a
\emph{post-condition}. Similarly to actions in the system meta-model,
pre- and post-conditions are represented as re-writing rules. However,
the subsystem represented in the pre- and post-conditions
is expressed by referring directly  to incident entities and their
connectivity and containment relationships. More precisely, a \emph{Condition} refers to a set of incident entities (\emph{IncidentEntityRef}). An \emph{IncidentEntityRef} is characterized by the \emph{name} of the \emph{IncidentEntity} it refers to, as well as some of the incident entities it contains (\emph{containment}). An \emph{IncidentEntityRef} can also refer to a set of connections (\emph{ConnectionRef}), where the referred incident entity is involved. A \emph{ConnectionRef} refers to a \emph{Connection} using its \emph{name}. In the rest of this section we describe how to use
the incident meta-model to
represent incident instances and patterns.

\subsubsection{Modeling Incident Instances}
\label{sec:model-incid-inst}

When the incident meta-model is used to represent an incident instance, a target, a resource, an activity initiator, and a location refer to specific system components and actors in a cyber-physical system. Thus, in an incident instance model, the \emph{name} and \emph{type} of incident entities refer respectively to the name (e.g., \emph{sl1}) and the class (e.g., \emph{SmartLight}) of a specific system component. Containment and connectivity relationships between incident entities can refer to containment and connectivity relationships between the corresponding system components in the cyber-physical system.

The incident instance model of the \emph{Research Center} incident is
depicted in Fig.~\ref{fig:incidentInstance}. We only show
details of two of the incident activities (``\emph{enter toilet}'' and
``\emph{connect laptop  to hvac}''). Activity ``enter toilet'' has a
\emph{visitor}  as an \emph{ActivityInitiator}, who has the
\emph{role} of an \emph{offender}. This Activity happens in room
\emph{hallway} (Location) and targets room \emph{toilet} (Target). As
shown in the model of the research centre in
Fig~\ref{fig:sysInstance}, the toilet also contains a smart light
(\emph{sl1}), which is connected physically to the bus network. Activity ``\emph{connect laptop to hvac}'' aims to establish digital connectivity between a laptop and an hvac. It is still  performed by the \emph{visitor} inside the \emph{toilet}. It targets the \emph{hvac} and uses the \emph{laptop} as a resource to establish connectivity.  As shown in the model of the research centre in Fig~\ref{fig:sysInstance}, the \emph{hvac} is connected physically to the \emph{bus network}. 

\begin{figure*}[t]
	\centering
	\includegraphics[width=\textwidth]{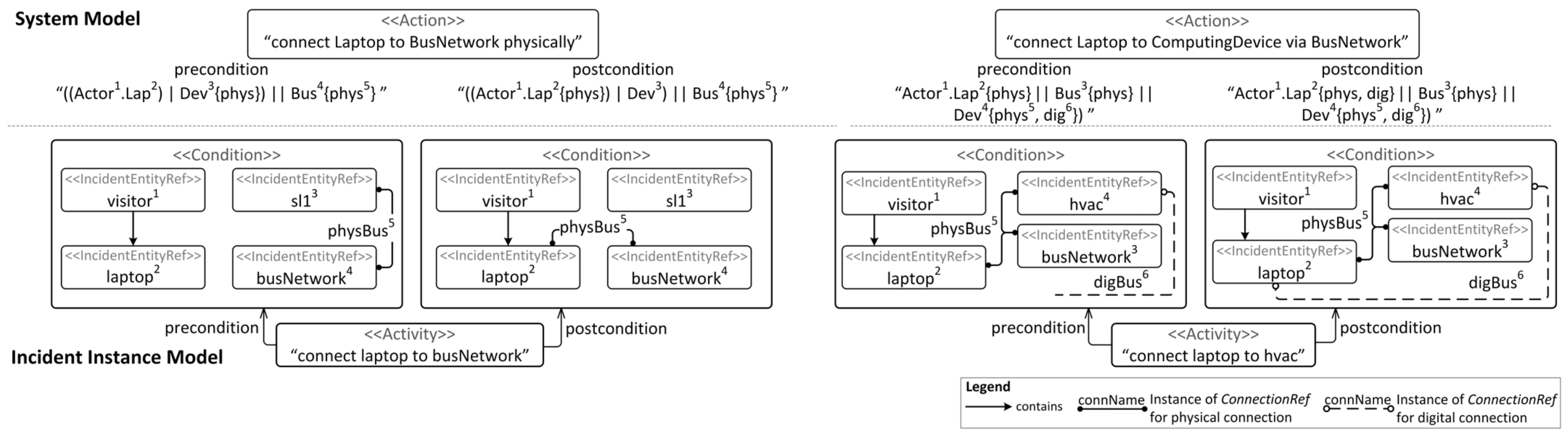}
	\caption{Mapping the system actions ``connect Laptop to BusNetwork physically" and ``connect ComputingDevice via BusNetwork" to the incident instance activities ``connect laptop to busNetwork" and "connect laptop to hvac" respectively.}\label{fig:actionMapping}
\end{figure*}

An activity in an incident instance has a direct mapping to an
action in the model of the cyber-physical system.  Pre- and
post-conditions of an activity are the same as pre-and
post-conditions of the corresponding action but component  types
are replaced with  concrete component instances from the
system. Containment, physical and digital connectivity relations
expressed in the pre- and post-conditions of an action are also
replaced with concrete relations expressed in the CPS model. For
example, as shown in the upper part of  Fig.~\ref{fig:actionMapping}, activity
``\emph{connect laptop to busNetwork}'' is associated with action
``\emph{connect Laptop to BusNetwork physically}'', which is
specified in Table~\ref{tab:actions}. $Room$, $Actor$, $Lap$, $Dev$ and $Bus$ 
are replaced by \emph{toilet}, \emph{visitor},\emph{laptop}, smart light \emph{sl1},
and \emph{busNetwork}, respectively.  Physical connectivity
$\{phys\}$ is replaced by \emph{physBus}, i.e. the physical
connectivity between the \emph{busNetwork} and \emph{sl1}. Activity
``\emph{connect laptop to hvac}'' is created based on action
``\emph{connect Laptop to ComputingDevice via BusNetwork}'' specified
in Table~\ref{tab:actions}. $Lap$, $Dev$ and $Bus$ are replaced by
\emph{laptop}, \emph{hvac}, and \emph{busNetwork},
respectively. Physical connectivity $\{phys\}$ is replaced by
\emph{physBus}, i.e. the physical connectivity between the \emph{busNetwork}
and the \emph{hvac}, while $\{dig\}$ is replaced by  a new digital
connectivity between the laptop and the hvac, which is created as an
effect of the action.
If the incident instance uses an activity that does not have any
corresponding action in the system model, it will be first necessary
to create a new action in the system model to which the activity can
be associated.

\begin{figure}[htbp]
	\centering
	\includegraphics[width=0.99\columnwidth]{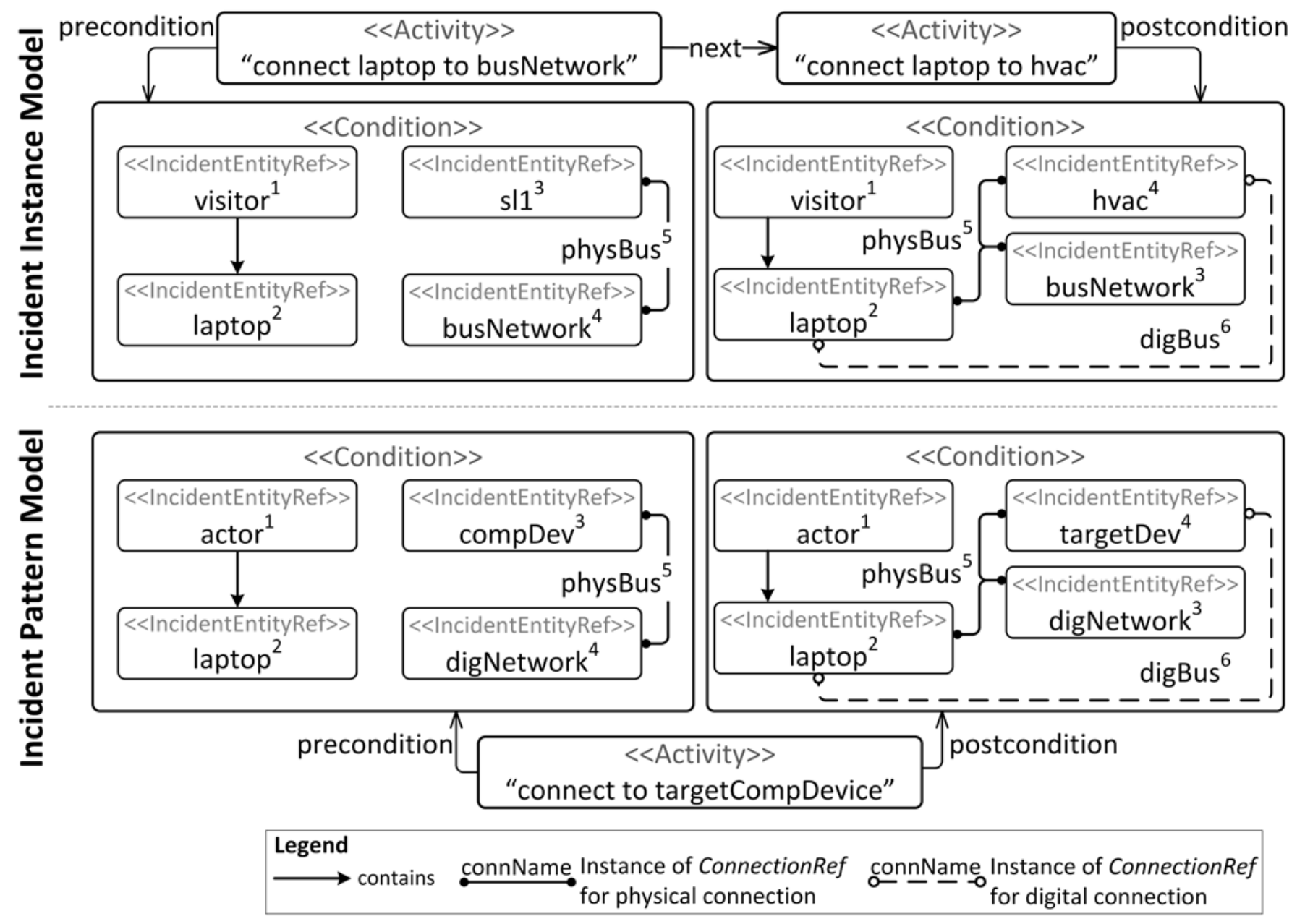}
	\caption{Merging the incident instance activities "connect laptop to busNetwork" and "connect laptop to hvac" into the incident pattern activity ``connect to targetCompDevice".}\label{fig:mergingActivities}
\end{figure}

\subsubsection{Modeling Incident Patterns}
\label{sec:model-incid-ptrn}

When an incident pattern is represented, a target, a resource, an
activity initiator, and a location can only refer to abstract system
components that can match more than one concrete component in a
cyber-physical system. Thus, in an incident pattern model the
\emph{name} of incident entities is just a random unique name (e.g.,
``compDev''), and the \emph{type} refers to an entity type  in the
smart building meta-model (e.g., \emph{ComputingDevice}).

A potential incident pattern model can be extracted from the incident
instance model shown in Fig.~\ref{fig:incidentInstance}. Such an
incident pattern model is shown in Fig.~\ref{fig:incidentPattern}.
Each Incident Entity refers to a more abstract entity (e.g., in Level
2 in the cyber-physical system meta-model). For instance, incident
instance \emph{sl1} which has the type \emph{SmartLight} is abstracted to an
incident entity that has type \emph{ComputingDevice}. Activities of an
incindent instance can also be merged. For example, the two activities
of the incident instance model, ``connect to busNetwork'' and
``connect to hvac'', are abstracted to one activity, ``connect to
targetCompDevice''. In this case,
the precondition is set to the precondition of the first activity
(``connect to busNetwork'') in the sequence. The postcondition is set to the
postcondition of the last activity (``connect to hvac'') in the
sequence. Pre- and post-condition of the new activity, ``connect to
targetCompDevice'' are shown in Fig.~\ref{fig:mergingActivities}. The process we follow to abstract incident
instances and merge a set of subsequent
activities is described in Section~\ref{sec:extraction}.

\section{Incident Pattern Extraction}
\label{sec:extraction}



In this section we describe the technique we propose to extract an incident pattern from an incident instance, automatically. Our technique includes two steps: merging and abstraction.

\begin{figure}[t]
	\centering
	\includegraphics[width=0.98\columnwidth]{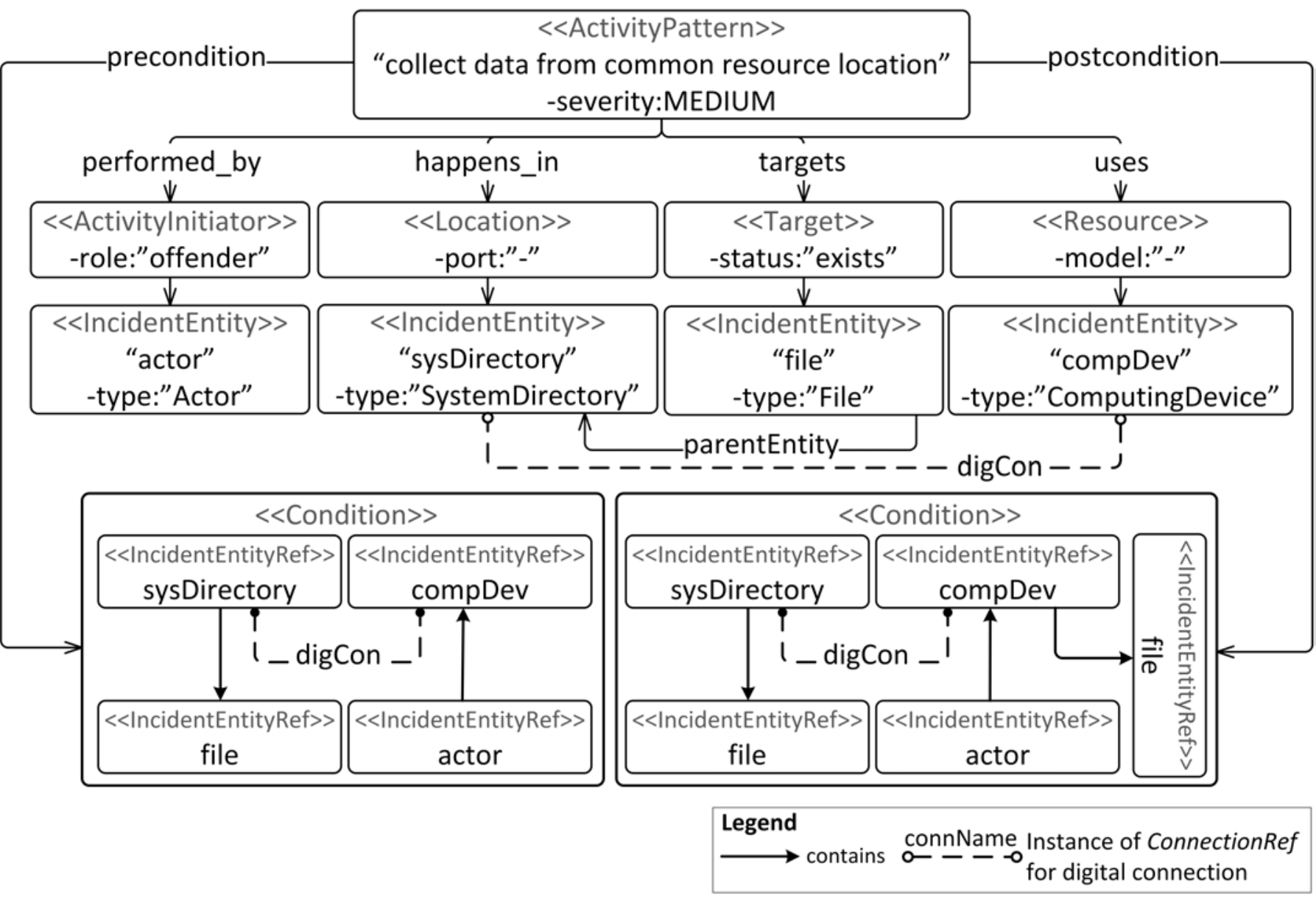}
	\caption{``Collect data from common resource location'' activity pattern.}\label{fig:colActPtr}
\end{figure}

\begin{enumerate}
	\item \emph{Merging:} We map a sequence of activities in the input incident instance to an \emph{activity pattern}, which represents an activity that is usually performed by an offender. In our approach, we defined  manually a set of activity patterns based on the Common Attack Pattern Enumeration and Classification (CAPEC) catalog~\cite{MITRECorporation}. CAPEC provides more than 500 common attack patterns. An attack pattern describes, textually, the common attributes and approaches that an offender can exploit to harm or weaken a target system. To create our activity patterns, we analyze the CAPEC attack patterns and  model them as activities of the incident meta-model. We only focused on attack patterns that could materialize in cyber-physical systems, particularly in smart buildings. Table~\ref{tab:actPatterns} shows the name, category, and abstraction level of the attack patterns that we used to model activity patterns. CAPEC abstraction levels include \emph{Standard}, which is a description of an attack technique (e.g., collect data from common resource locations~\footnote{https://capec.mitre.org/data/definitions/150.html}), and \emph{Meta}, which is an abstract characterization of an attack technique (e.g., Excavation\footnote{https://capec.mitre.org/data/definitions/116.html}).

     \begin{footnotesize}
        \begin{table}[htbp]
          \centering
          \caption{Modeled CAPEC attack patterns.}
          \label{tab:actPatterns}
          \begin{tabular}{|p{2.8cm} | p{2.8cm} | p{1cm} |}
            \hline
            \textbf{Name} & \textbf{Category} & \textbf{Level} \\ [0.1cm] \hline 
		
            Collect Data from Common Resource Locations & Collect and Analyze Information & Standard \\ [0.45cm]\hline  
		
            Sniffing Attacks & Collect and Analyze Information & Standard \\  [0.45cm] \hline 
		
            Content Spoofing & Engage in Deceptive Interactions & Meta \\ [0.45cm] \hline 
		
            Establish Rogue Location & Engage in Deceptive Interactions & Standard \\  [0.45cm] \hline
		
            Exploitation of Trusted Credentials & Subvert Access Control & Meta \\ [0.45cm]\hline  
		
            Physical Theft & Subvert Access Control & Meta \\  [0.45cm] \hline 
		
            Using Malicious Files & Subvert Access Control & Standard \\ [0.45cm]\hline 
		
            Functionality Bypass & Abuse Existing Functionality & Meta \\  [0.45cm]\hline
		
            Email Injection & Inject Unexpected Items & Standard \\[0.45cm]\hline  
		
            Hardware Integrity Attack & Manipulate System Resources & Meta \\\hline 
		
            \hline  
          \end{tabular}
	
        \end{table}
         \end{footnotesize}

A CAPEC attack pattern is characterized by a \emph{description}, an indication of its \emph{severity}, a set of  \emph{resources} that offenders require to perform the attack, some \emph{prerequisites} that should  be met, and some \emph{related weaknesses} that must be present in the target system ---at least one of them--- to perform the  attack successfully. Weaknesses are usually represented using ids of relevant CWE (Common Weakness Enumeration) vulnerabilities. For example, Table~\ref{tab:CAPECAttackPatterns} shows attack pattern ``Collect Data from Common Resource Locations''.

          \begin{center}
            \begin{footnotesize}
              \begin{table}[htbp]
                \centering
                \caption{Collect Data from Common Resource Locations.}
                \label{tab:CAPECAttackPatterns}
                \begin{tabular}{|p{7.5cm}|} 
                  \hline      
                  \textbf{Description} \\\hline
                  An adversary exploits well-known locations for resources for the purposes of undermining the security of the target. Even when the precise location of a targeted resource may not be known, naming conventions may indicate a small area of the target machine's file tree where the resources are typically located. For example, configuration files are normally stored in the /etc director on Unix systems. Adversaries can take advantage of this to commit other types of attacks. \\\hline
                  \textbf{Severity} \\\hline
                  Medium \\\hline
                  \textbf{Prerequisites} \\\hline
                  The targeted applications must either expect files to be located at a specific location or, if the location of the files can be configured by the user, the user either failed to move the files from the default location or placed them in a conventional location for files of the given type. \\\hline
                  \textbf{Resources Required} \\\hline
                  None \\\hline
                  \textbf{Related Weaknesses} \\\hline
                  CWE-ID: 552. Files or Directories Accessible to External Parties \\\hline
      
                \end{tabular}

              \end{table}
            \end{footnotesize}
          \end{center}

		
	The activity we created to represent attack pattern ``Collect Data from Common Resource Locations'' is shown in Fig.~\ref{fig:colActPtr}. To model attack patterns we first read and examine the \emph{description} of the attack pattern to identify some keywords that allow us to define the activity initiator, location, target and resource. For example, keyword   ``adversary'' indicates that the activity is performed by  an \emph{Actor} playing the role of an \emph{offender}. Keyword ``target'' refers to the activity target, which in this case is a \emph{File} inside a \emph{SystemDirectory}. \emph{Resource}s can be identified from the the \emph{resources required} and/or the \emph{description} of the attack pattern. In this example, although no resources are necessary, we still considered important for an attacker to use a computing device in order to access a file. Thus, in our activity pattern we define a \emph{Resource} to be a \emph{ComputingDevice}. Relationships between entities can also be identified from the \emph{description} of the attack pattern. For example, we represent a containment relationship between the \emph{File} and the \emph{SystemDirectory}. 

	Second, pre- and post-condition of the activity pattern are identified by examining the \emph{prerequisites}, \emph{weaknesses}, and \emph{description} of the attack pattern. The \emph{prerequisites} and \emph{weaknesses} allow us to identify the precondition. In this example, from the \emph{prerequisites}, we identified that the targeted \emph{File} should be \emph{contained} in the \emph{SystemDirectory}. From the \emph{weaknesses} we identified, as part of the precondition, that the offender should have accessibility to the \emph{Location}, which we translated as connectivity between the offender's \emph{ComputingDevice} and the \emph{SystemDirectory}. 
	The post-condition is inferred from the \emph{description}. In this example, we identified that the offender's \emph{ComputingDevice} should contain the targeted \emph{File}. Finally, the \emph{severity} of the attack pattern is assigned directly to severity of the activity pattern. In this case, the \emph{severity} of the activity pattern is set to \emph{MEDIUM}. 


	
	

In addition to CAPEC attack patterns, we created pattern activities based on common actions that occur in a smart building such as movement between rooms, and connectivity to a network. For example, we created an activity pattern that expresses the activity of digitally connecting one's computing device (e.g., laptop) to another computing device (e.g., HVAC) in a system. The created activity pattern is shown in Fig.~\ref{fig:conActPtr}. This activity pattern allowed the merging of the two activities shown in Fig.~\ref{fig:mergingActivities}.

Our extraction technique identifies sequences of activities in an incident instance that satisfy the pre- and post-condition of the activity patterns. A pre- or post-condition in an activity pattern matches a pre- or post-condition of an activity in the incident instance if the entity types (classes) referred in the condition of the activity pattern are (super) types of the concrete system components referred in the incident instance. Also the pre- and post-conditions in the activity of the incident instance should support the same containment and connectivity relationships expressed in the conditions of the activity pattern. In other words, pre- or post-condition of an activity pattern is converted into a Bigraph, which is then matched against the pre- or post-condition of an activity, which is also represented as a Bigraph. If a pre-condition of an activity pattern is matched, then we try to find a post-condition of an activity in the incident instance that matches the post-condition of the pattern. If a match for the post-condition of the activity pattern is found, then the activity sequence, which begins from the activity that matched the pre-condition of the activity pattern and ends with the activity that matched the post-condition of the pattern, is preserved for when we determine which set of activity patterns to use for merging. 

Because different activity patterns can match an overlapping set of activities in an incident instance, we propose a strategy to prioritize the activity patterns to be selected for the matching. More precisely, we aim to maximize the number of activity patterns that match a non overlapping sequence of activities in an incident instance and that have a maximum severity. We implemented the matching as a constraint solving problem, using Choco~\cite{Prudhomme}. The sequences of activities in the incident instance matching activity patterns are replaced by new activities. For each activity sequence we create a new activity that has as a pre-condition the pre-condition of the first activity in the sequence, and as a post-condition the post-condition of the last activity in the sequence. Unmatched activities will remain in the incident instance sequence, however, some of their attributes can be abstracted such as target and location.

	\item \emph{Abstraction:} After merging, we replace each component referred to in an incident actor, target, location or resource with a more abstract representation. We define rules of thumb to decide the level of abstraction for each system component. Generally, we abstract a component's type to one level up (i.e. to a more general type) in the system meta-model. For example, a component with type \emph{SmartLight}, level-3, is abstracted to \emph{ComputingDevice}, level-2. Similarly, we abstract a component's \emph{Connection}s. For example, a \emph{BusConnection} can be abstracted to a \emph{DigitalConnection} following the one-level-up rule. In other cases, instead, we keep the same level of abstraction. For example a specific room (e.g., \emph{hallway}) will be simply referred to as \emph{Room}. Determining the appropriate level of abstraction for different types is a challenging task and there is no silver bullet solution. Our rules were derived from our experience in using our incident meta-model to represent various incidents in smart buildings. Incident patterns can also be reviewed by a user to adjust abstraction level, if needed. These adjustments can be used to inform future abstractions. Moreover, in the abstraction process of a component, its \emph{name} is changed (e.g.,``toilet'' becomes ``room1''). Only selected \emph{properties} of a component are included in the abstract representation. Currently, property selection is done manually; i.e. a user determines which properties can be in the abstract version.

The Abstraction levels can be useful in determining which activity patterns to apply when extracting an incident pattern. For example, if more abstraction in an incident pattern is required, then we can apply activity patterns with \emph{Meta} abstraction level instead of \emph{Standard}. This can be done, for example, by adding a criterion to the constraint problem solver. This criterion can indicate that a solution (i.e. set of selected activity patterns) should contain more \emph{Meta} activity patterns than the previous solution.

Finally, beside abstracting the activities and entities, the technique abstracts information provided in the \emph{CrimeScript}. In particular, the script \emph{category} is changed from INSTANCE to PATTERN.
\end{enumerate}

\begin{figure}[t]
	\centering
	\includegraphics[width=0.98\columnwidth]{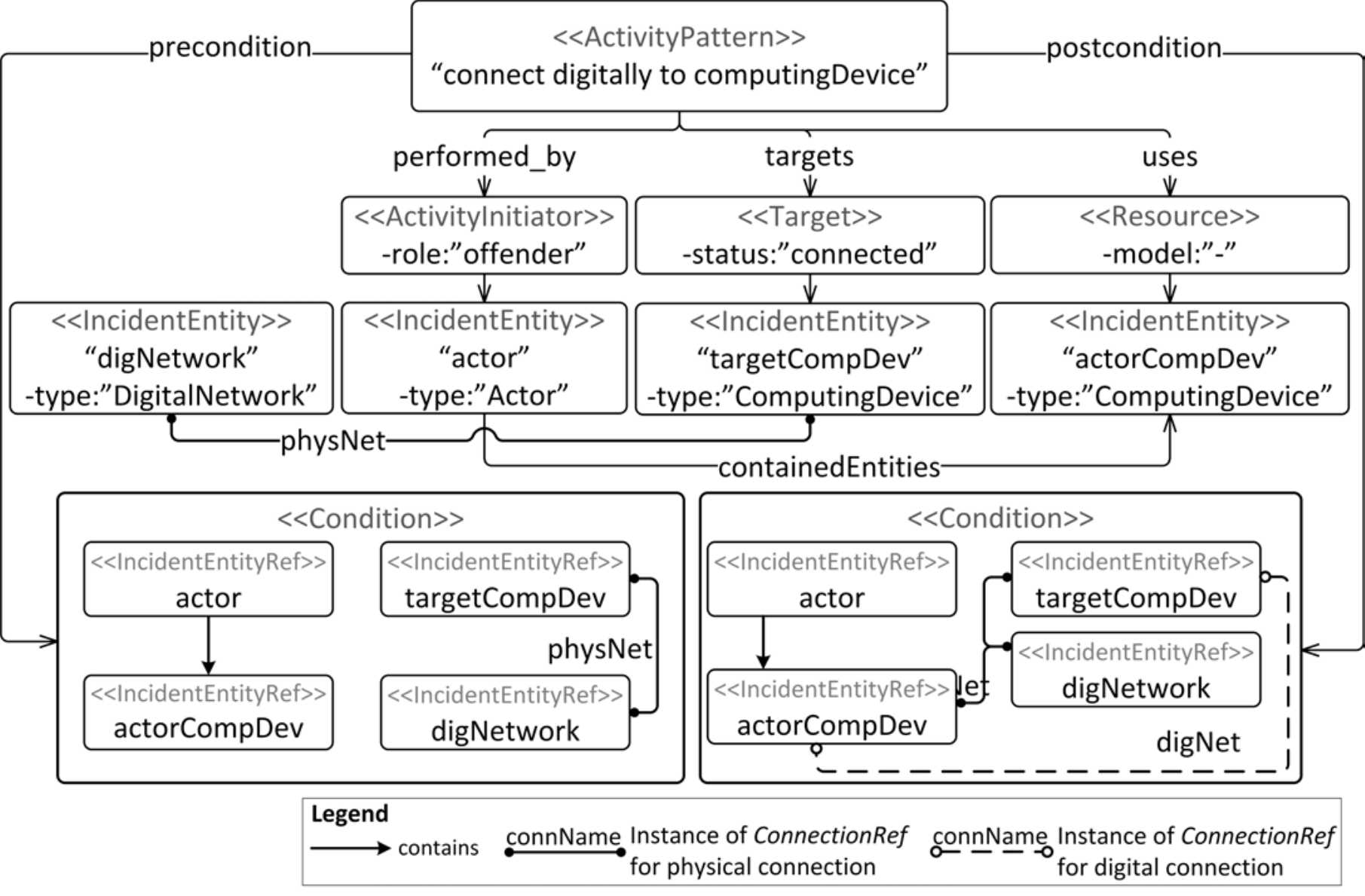}
	\caption{``connect digitally to computingDevice'' activity pattern.}\label{fig:conActPtr}
\end{figure}


\section{incident pattern instantiation}
\label{sec:instantiation}

We present our automated instantiation technique to facilitate the assessment of whether and how prior incident instances can re-occur or manifest in cyber-physical systems. Assessment involves determining if a CPS can satisfy an incident pattern, which is extracted from incident instances. If so, then the technique identifies all system traces; i.e. sequences of actions that satisfy the incident pattern activities.

Instantiation involves identifying components in a system that match to incident pattern entities, and then identifying sequences of actions in the system that satisfy the pattern sequence of activities. 


To identify system components that match to an incident pattern entity, we define a matching criteria as follows. First, matching \emph{type} of an entity. A system component is a potential match to an incident pattern entity if the component has a type that is the same or less abstract than that of the entity type. In other words, the system component should have the same class as the entity or any of its subclasses. For example, the incident pattern entity in Fig.~\ref{fig:incidentPattern}, which has the type \emph{ComputingDevice}, has the following potential matching components in the research center: \emph{SL1}, \emph{FireAlarm}, and \emph{Server}. Second, matching the type of an entity's parent. Similar to the first criterion, the type of a system component's parent should be the same or less abstract than that of an entity’s parent. Third, matching the number and type of contained entities. In an incident pattern, the \emph{knowledge} about contained entities can be specified by EXACT or PARTIAL. EXACT implies that a system component should contain a number of components that is equal to that of an entity in the incident pattern. PARTIAL implies that a system component should have a number of components that is more than or equal to that of an entity. Each of the contained entities should be matched to a contained component. If \emph{knowledge} is EXACT, then all contained components should have a match to a system component. Fourth, matching the number and type of connections. Similar to contained entities, connections are matched against their number and types depending on the knowledge (i.e. EXACT or PARTIAL). Fifth, matching attributes. If an attribute is defined in an entity, then a matching component should possess same attribute. For example, if the attribute \emph{status} of an entity is defined to be ``OPEN", then all matched components should have ``OPEN" as a status.

If an entity has no matches, this implies that the system does not contain components that match the specified criteria for this entity. Therefore, the incident pattern cannot be instantiated in the system. If all entities have matches, then the technique generates all possible sets of system components that map to the set of incident pattern entities. For example, an entity of type \emph{ComputingDevice} can be mapped to \emph{Research Center} components \emph{{SL1, FireAlarm, Workstation}}. An entity of type \emph{Actor} can be mapped to \emph{{Visitor}} in the \emph{Research Centre}. Possible component sets for the two entities \emph{{ComputingDevice, Actor}} include \emph{{{SL1, Visitor}, {FireAlarm, Visitor}, and {Workstation, Visitor}}}. Possible sets can be restricted by maintaining relationships found between incident pattern entities. For example, if the entities (\emph{ComputingDevice} and \emph{Actor}) are contained by the same entity (e.g., \emph{Room}), then all generated sets should respect this relationship. If \emph{{FireAlarm, Visitor}} set does not satisfy this containment relationship, then it will be removed from the possible sets of components.

For each set of system components the technique identifies states that satisfy the incident pattern activities (preconditions \& postconditions). To do so, we feed the technique a Labelled Transitions System (LTS). This LTS is based on the system instance model, and is generated using BigraphER. Then, the technique identifies all states that satisfy the precondition and postcondition of each activity. Satisfaction of a pre-/post-condition is carried out using Bigraph matching, in which a condition is converted into a Bigraph, then matched against a system state. We use \emph{LibBig}~\cite{Miculan2014} library to perform Bigraph matching.


After identifying all states that satisfy the conditions of the incident pattern activities, the technique identifies traces. The technique finds all possible traces, in which each trace contains a sequence of states that satisfy all the activities. The first state satisfies the precondition of the first activity, and the last state satisfies the postcondition of the last activity. The technique finds a trace by using a Breadth First Search (BFS) algorithm. The algorithm identifies all non-cyclic transitions between two states in an LTS. The technique may identify a large number of traces that satisfy the incident pattern activities. Therefore, to reduce the number of identified traces, the technique performs the following. It skips traces that contain two states (e.g., A \& B) that satisfy the same condition, and state (A) has a trace to state (B). This is done since traces from (B) will be identified by BFS independently from (A). The technique also bounds trace's length. 

\begin{figure}[tbhp]
	\centering
	\includegraphics[width=0.99\columnwidth]{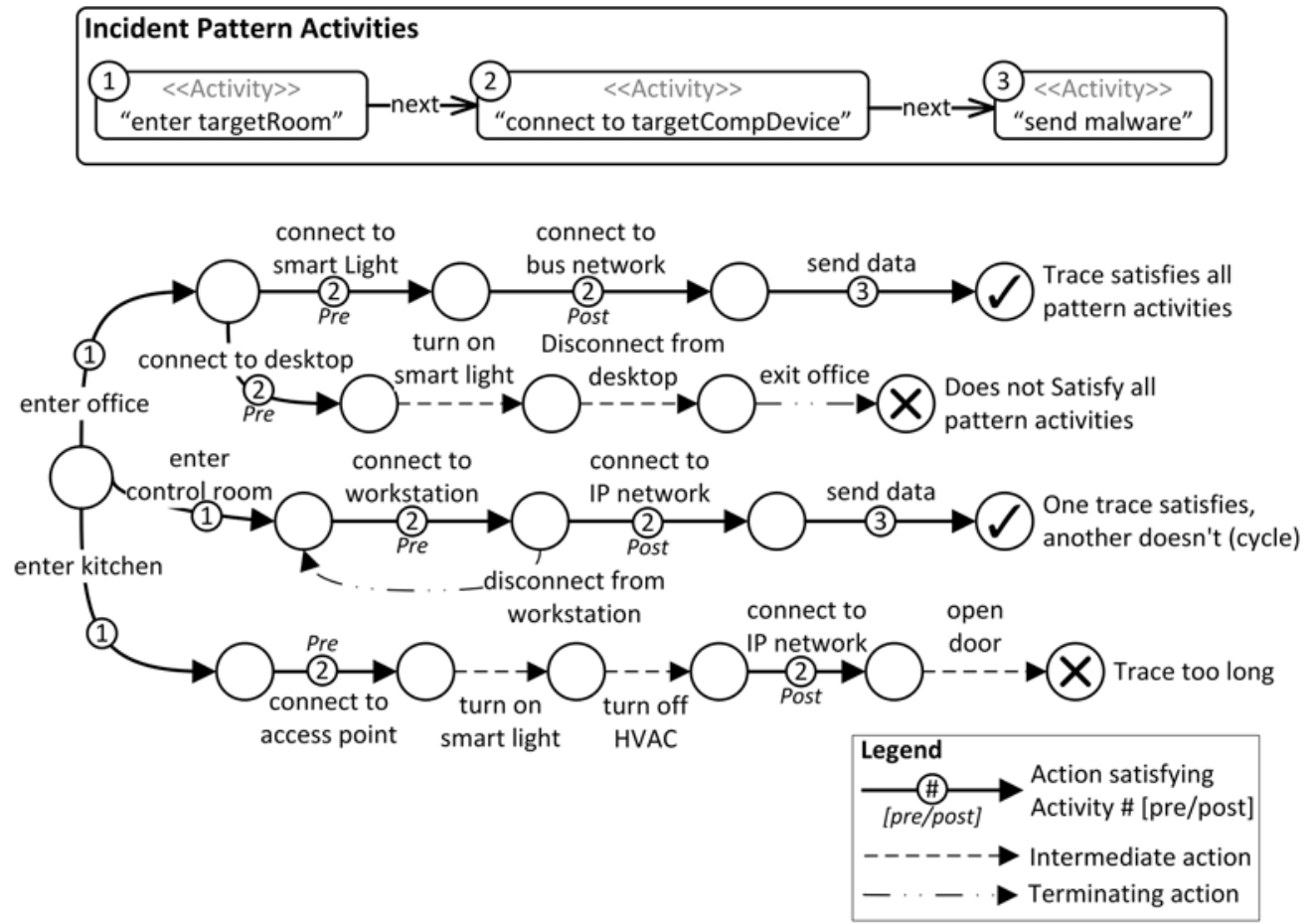}
	\caption{An instantiation of incident pattern activities in a LTS.}
	\label{fig:LTS}
\end{figure}

An instantiation of the incident pattern activities, which are shown in Fig.~\ref{fig:incidentPattern}, in an LTS is shown in Fig.~\ref{fig:LTS}. The instantiation identifies two correct traces that satisfy all activities. An action that satisfies an activity is denoted by a solid arrow. The other traces are not identified as correct traces for various reasons. First, one of the traces does not satisfy all the activities (i.e. ``send malware" is not satisfied). Second, a trace forms a cycle (i.e. ``connect to workstation" \& ``disconnect from workstation" actions). An action can be a terminating one if it is the last action in a trace and it does not satisfy the last activity in the incident pattern, or if it forms a cycle. Third, one trace is longer than the maximum number of actions (6 in this example) allowed for a correct trace. A trace might contain intermediate actions i.e. actions that are irrelevant to the satisfaction of an incident pattern such as ``turn on smart light".
 
A post-instantiation analysis can be carried out. The analysis identifies shortest traces, actions occurrences in transitions, and shortest traces with specific action occurrence percentage (i.e. the number of times an action appeared in the traces to the total number of traces). Shortest traces can provide some insight about the minimum actions that can lead to an incident in a cyber-physical system. A shortest trace would contain at least a number of actions that is equal to the number of activities in an incident pattern. A user could be interested in knowing all traces that have actions with a specific occurrence percentage (e.g., \textgreater= 90\%). This can be useful in identifying system actions that are most relevant to an incident pattern.


\section{Evaluation}
\label{sec:evaluation}

We evaluated the two automated techniques, incident pattern \emph{extraction} and \emph{instantiation} via two case studies. Both case studies are inspired by real-world systems and incidents. We evaluated three aspects, which are:

\emph{Correctness}. Does the output of each technique adhere to our meta-models and system dynamics? In other words, entities and relationships found in output should correspond to entities and relationships in the meta-models, and also to the Labeled Transition System (LTS). In the case of incident pattern extraction, this means that the generated incident pattern should contain only relationships and entities that exist in the system and incident meta-models. For example, a \emph{SmartLight} should be abstracted to one of the classes defined in the abstraction levels for it such as \emph{ComputingDevice}. In addition, the generated incident pattern should, if instantiated in the same system in which the incident instance occurred, regenerate the same activities found in the incident instance model. In the case of instantiation, this means that generated traces should correspond to traces that exist in the LTS.

\emph{Scalability}. Can the techniques generate output for larger systems? This, in particular, is relevant to the instantiation technique since it requires identifying all possible transitions in a LTS.

\emph{Performance}. How long does it take a technique to generate an output? For instantiation technique, this means the amount of time it requires to identify system states that satisfy all activities of a given incident pattern.

\subsection{Experimental Setup}
Our evaluation was based on a real floor layout design of an organization (here, referred to as RC1), which we had access to. We modified the floor design by adding various smart devices (e.g., smart lights, motion sensors, air conditioning units) to mimic real smart building setups. After that, we modeled RC1 using our smart building meta-model. We then analyzed the dynamics of RC1 using Bigrapher software tool to obtain an LTS. We generated various sizes of states and transitions (10,000 [33,850] to 100,000 [445,028] states [transitions]). Similarly, we modeled and analyzed another organization (here, referred to as RC2). We developed two incident scenarios. First scenario involves an incident instance that occurred in RC1. The incident instance is about collecting data transmitted over internal bus network. It consists of six activities, which are: enter RC1, move to a room containing access point, connect to access point, connect to bus network, collect data transmitted, and analyze collected data. Second scenario involves an incident pattern that is instantiated in RC2. The incident pattern consists of three abstract activities, which are: reach a location in smart building that contains a device connected to internal network, gain access to the internal network via device, and collect transmitted data. Techniques are tested on a virtual machine that has Ubuntu 18.04.1 LTS 64bits as operating system, Intel® Xeon® CPU E5-2660 2.2GHz (32 CPUs), and 64GB of memory. 

\subsection{Results}
We evaluated the \textit{correctness} of the extraction technique via the first scenario, in which the technique generated an incident pattern out of an incident instance that occurred in RC1. Evaluation was carried out by examining and verifying that all generated entities and activities conforms to the meta-models. In other words, generated abstract activities should conform to activity description in the incident meta-model and also to activity patterns, if any is used. In addition, abstracted entities should conform to the abstraction levels defined in the system meta-model. We also evaluated correctness via instantiating the generated incident pattern in the same system (i.e. RC1), and verified that it can produce the original incident instance. We evaluated the correctness of the instantiation technique by examining and verifying that all generated traces contained minimum actions for mapping the incident pattern into RC2 in the second scenario. The minimum actions were identified manually for the incident pattern. We also verified that all generated traces were valid; i.e. they can be identified in LTS of RC2. 

We evaluated the \textit{scalability} of the extraction technique by modifying the first scenario to include more assets and actions, and then generated an incident pattern. The technique was able to generate correct incident pattern. For the instantiation technique, we applied the second scenario over various LTS sizes; i.e. we instantiated an incident pattern in RC2 over LTS sizes between 10,000-100,000 states. Table~\ref{tab:InstantiationOutput} shows the number of generated traces for different LTS sizes for RC2. As can be noticed from Table~\ref{tab:InstantiationOutput}, the number of generated traces increased as LTS size increased. This can be due to several reasons. One reason is that there are new traces in the LTS that satisfy the incident pattern. Another reason is the existence of traces containing the same actions that satisfy the incident pattern, but they also contain different \emph{irrelevant} actions; i.e. actions that are not related to an incident. 

To reduce traces that contain irrelevant actions, we performed filtering over generated traces. Filtering is done by mining frequent sequential patterns using the ClaSP Algorithm~\cite{Gomariz2013}. We define frequent sequential patterns as the shortest traces that share a maximum length partial trace. We used an open source data mining library called SPMF\footnote{http://www.philippe-fournier-viger.com/spmf/}, which provides an implementation of the ClaSP algorithm. We filtered the generated traces as follows. We created a functionality that reads the generated traces. Then finds all traces that have the shortest sequence length. Following this, it converts shortest traces into a format compatible with SPMF. The functionality then applies the ClaSP algorithm over converted traces. Finally, it stores results (i.e. filtered traces) given by the algorithm. Filtered traces are shown in Table~\ref{tab:InstantiationOutput}. We can notice a significant reduction in the number of traces between \emph{Generated} and \emph{Filtered}. This can be useful since one can identify minimum actions in the system that can lead to an incident. Other criteria for filtering can also be applied. For example, filtering traces with an occurrence percentage higher than a certain value.

We define a false positive as a trace that is generated but does not satisfy the incident pattern. In this case, since all traces must satisfy the incident pattern to be generated, then there cannot be false positives. However, It can be the case that a generated trace that satisfies the incident pattern is deemed to be false positive. In this case, a false positive is defined as a generated trace that does not actually form a potential incident in the system. For this to be done, a user needs to manually inspect a trace to determine its potential as an incident. We define a false negative as a trace in the LTS that satisfies the incident pattern but is not generated. False negatives can exist in our approach in one case, that is if a trace is longer that the set bound for identifying traces. In other words, limiting the search for traces to a certain length implies that any trace that satisfies the incident pattern, but has a length over the limit will not be generated. This can be mitigated by adjusting the length. If needed, the length can be adjusted to the maximum trace length that a trace can have in the LTS. However, this will impose challenges for the scalability and performance of the approach.  

\begin{figure}[b]
	\centering
	\includegraphics[width=0.99\columnwidth]{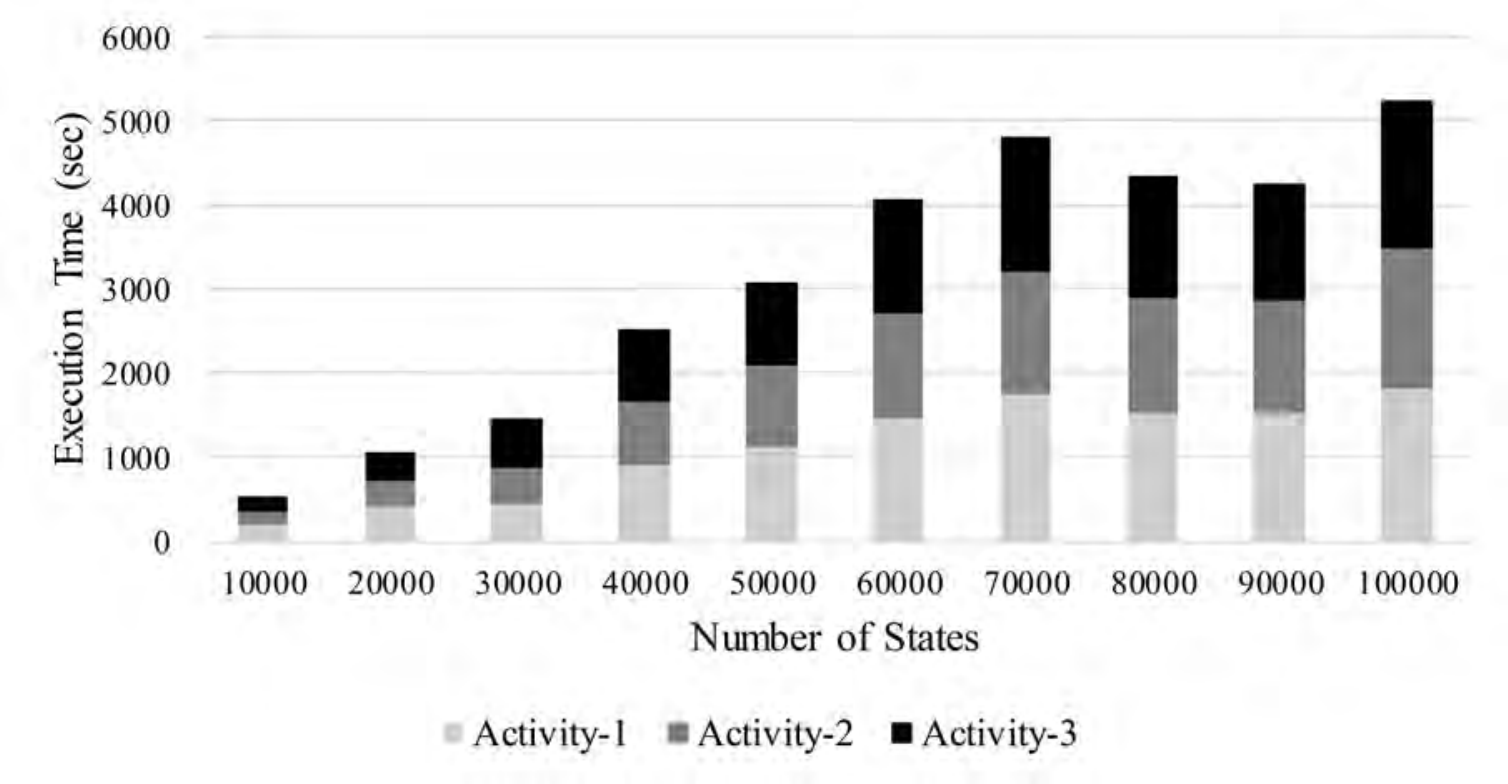}
	\caption{Execution time of incident pattern activities measured over various LTS sizes (10K-100K states) of RC2.}
	\label{fig:execution}
\end{figure}

\begin{figure}[htbp]
	\centering
	\includegraphics[width=0.99\columnwidth]{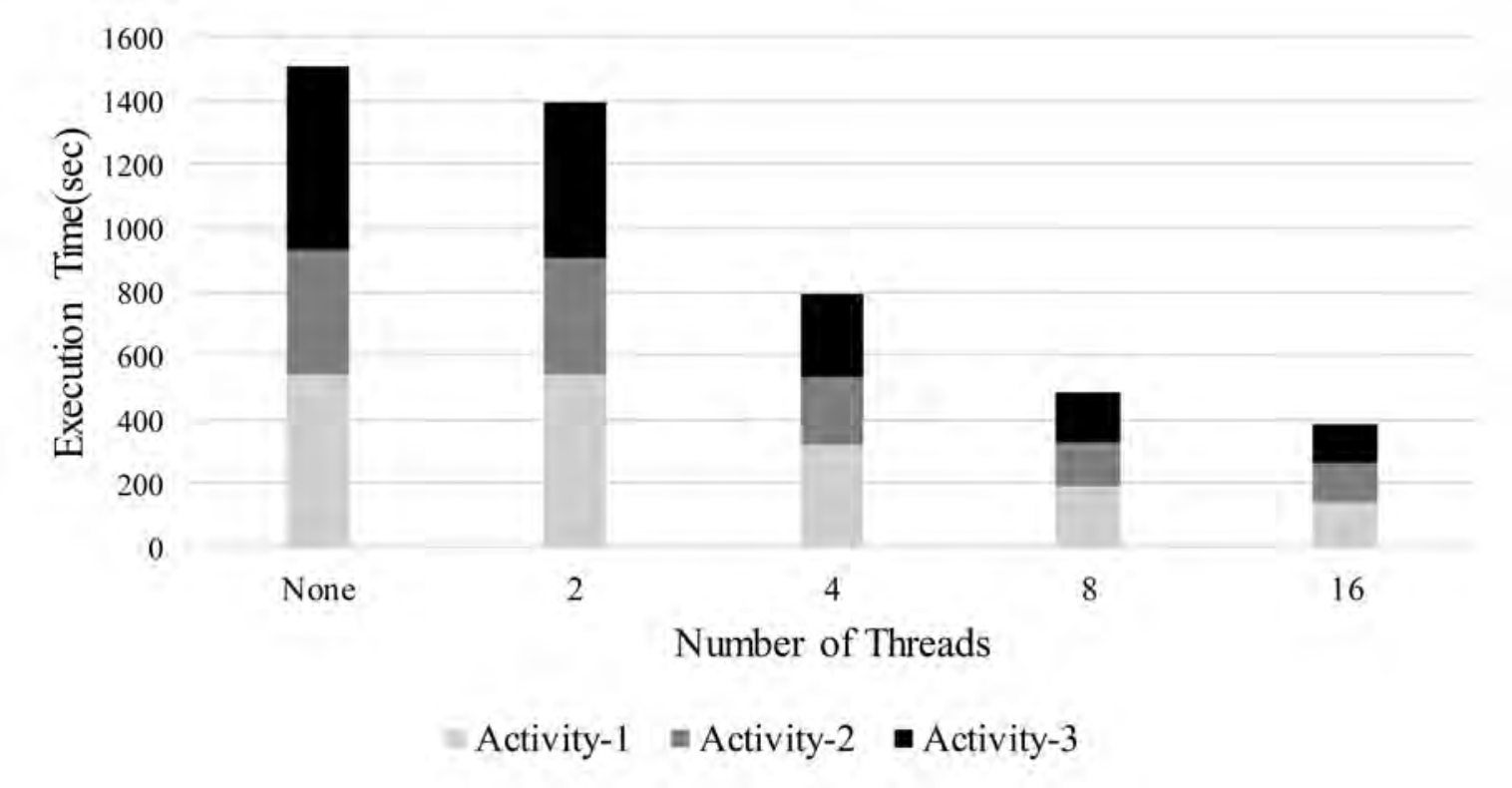}
	\caption{Execution time of instantiating incident pattern activities using different number of threads in RC2 with LTS of size 10,000.}
	\label{fig:threading}
\end{figure}

We evaluated \textit{performance} by measuring execution time. Execution time means the time that the instantiation technique required to identify all states that satisfy an incident pattern activities. We measured the execution time of the instantiation technique using different LTS sizes. The execution time for instantiating the incident pattern activities of the second scenario over different LTS sizes is depicted in Fig.~\ref{fig:execution}. Execution time varied depending on the LTS size. To enhance performance, we implemented parallelism in various parts of the instantiation technique. For example, matching system states to activity conditions is carried out by dividing all states into subsets that are then handled by different threads. Fig.~\ref{fig:threading} shows performance using different number of threads. As can be noticed from Fig.~\ref{fig:threading}, performance improved by using parallelism. Execution time was almost halved using 4 threads, and reached almost quarter the original time (i.e. no threading) using 16 threads. 

\begin{table}[t]
	\centering
	\caption{Output of the instantiation technique applied over different LTS sizes of RC2. Output shown is all generated  traces and relevant ones.
		\label{tab:InstantiationOutput}}
	\begin{tabular}{cccc} 
		
		\hline 
		\multicolumn{2}{l}{\textbf{~~~~~~~~LTS}} &&\\  \cline{0-1} 
		
		\textbf{States}  & \textbf{Transitions} & \textbf{Generated traces} & \textbf{Filtered traces}  \\ 
		\hline
		10,000  & 33,850      & 720                         & 40                   \\ [0.1cm]
		20,000  & 73,734      & 2,845                       & 90                   \\ [0.1cm]
		30,000  & 110,569     & 5,403                       & 120                   \\ [0.1cm]
		40,000  & 158,477     & 10,648                      & 160                   \\ [0.1cm]
		50,000  & 198,771     & 14,777                      & 200                   \\ [0.1cm]
		60,000  & 252,897     & 23,848                      & 240                   \\ [0.1cm]
		70,000  & 295,160     & 98,720                      & 265                   \\ [0.1cm]
		80,000  & 349,517     & 143,186                     & 305                   \\ [0.1cm]
		90,000  & 399,319     & 184,269                     & 310                    \\ [0.1cm]
		100,000 & 445,028     & 216,561                     & 340                   \\ 
		\hline
	\end{tabular}

\end{table}

\subsection{Threats to Validity}

Our results suggest that our techniques were able to produce correct output for the scenarios in which they were tested in. However, there are several threats, internal and external, that may affect the interpretation and generalizability of our results. 

One internal threat is that matching states to conditions could be wrong. We use \emph{BigraphER} to generate Bigraph states, but use \emph{LibBig} to match generated Bigraph states to Bigraph conditions. This may lead to incorrect mis/matching since different implementations of Bigraphical Reaction Systems (BRS) may have different interpretation of BRS. To eliminate this, we verified correctness of the matching by comparing matching results obtained from \emph{BigraphER} and that from \emph{LibBig} library. Another threat is that the implementation of both techniques can contain bugs that may affect the output. We made sure that this is minimized by testing each functionality used by the techniques. Furthermore, we used known algorithms in different parts of the techniques. For example, we used shortest path algorithm to identify transitions in the instantiation technique. We also used constraint problem solver to identify an optimized solution in the extraction technique. A threat that is specific to the generation of trances in the instantiation technique is bounding. Bounding trace identification to certain length can lead to missing potential traces that satisfy an incident pattern. This can be mitigated by defining an adaptive bounding with a minimum value. Bounding can change based on the number of actions in an LTS, and also number of activities in an incident pattern. The minimum is defined to be equal to the number of activities in an incident pattern.

An external threat is the usefulness of generated incident knowledge i.e. incident patterns. An organisation may decide that an incident pattern is not abstract enough to be shared, or it is too abstract or too detailed to be instantiated. To tackle this, we proposed the use of CAPEC to assist in creating incident patterns. Moreover, we intend to assess the practical applicability of the approach to evaluate usefulness. Another threat is the modelling of systems and incidents, which may not correspond to real systems and incidents. To mitigate this, system and incident models, which are used in the case studies, are inspired from real systems and incidents. Although we evaluated the techniques over a single floor in RC1 and RC2, for larger systems, scalability may become an issue. This limitation can be addressed by decomposing a smart building into several parts (or floors) that can be then analyzed. For this, BRS is a suitable choice since it is capable of decomposing large systems and then analyzing them. Our techniques can then reason about the different parts.

\section{Related Work}
\label{sec:relatedWork}

A number of related approaches have been proposed for sharing incident information. Most of these approaches focus on representing and analyzing cyber incidents. Few approaches try to represent and analyze incidents in CPSs, and fewer provide a way for sharing incident knowledge between CPSs. 

Current attack modeling techniques (e.g., attack graphs~\cite{Lallie2017}) focus on representing and analyzing how a traditional cyber attack (e.g., denial of service) can occur. As these techniques do not account for the interactions between cyber and physical components, they may not be suitable to represent and analyze cyber-physical incidents~\cite{yampolskiy2015}. Existing resources for incident information are also focused on cyber incidents. For example, the Common Vulnerabilities \& Exposures (CVE)~\cite{MITRECorporation} is a publically available dictionary of known cybersecurity vulnerabilities in software and devices. 

Other work, such as~\cite{Liu2013}, focus on analyzing specific attacks (e.g., switching attacks) that can occur in certain application domain of CPSs, such as smart grids. Thus, developed techniques cannot be applied to analyze different types of attacks that can also happen in other application domains. Additionally, they do not represent incident entities (e.g., motivation) that can be useful in case an investigation is required. Hawrylak et al.~\cite{Hawrylak2012} propose a Hybrid Attack Graph (HAG) to model cyber-physical attacks. Their approach produces a graph of all possible ways a set of exploit patterns can be applied to a system. However, the approach focuses on representing malicious actions that exploit vulnerabilities found in some devices and does not consider other non-malicious interactions between cyber and physical components that can lead to undesired state in a system.

Few approaches tackle security incident representation and sharing. Boll{\'{e}} and Casey~\cite{Bolle2018} proposed an approach to identify and share linkages between cyber crime cases in order to support digital investigations. The approach employs exact and near string similarity algorithms (e.g., Levenshtein algorithm) to identify similar digital traces between different cyber crimes. Their current implementation focuses on identifying similarities in email addresses. While it can be useful to identify and share similar digital traces, the approach focuses on cyber interactions, which may not be sufficient for investigating incidents in CPSs. STIX (Structured Threat Information Expression)~\cite{Mitre2012} has been proposed as a language for representing Cyber Threat Intelligence (CTI). CTI is shared using TAXII (Trusted Automated Exchange of Intelligence Information)~\cite{OASISOpen}. STIX provides only a representation for CTI without any analysis of how CTI should be extracted from one organization or how it should be instantiated in another. Fani and Bagheri~\cite{Fani2015s} proposed a light-weight security incident ontology to represent security events. The ontology consists of temporal, spatial, and event entities. However, their ontology is centralized around events and do not represent other entities that can be relevant for an investigator such as motive.       

\section{Conclusions \& Future Work}
\label{sec:conclusion}

In this paper we proposed a novel approach for representing and sharing security incident knowledge across different organizations. We suggested that knowledge of security incidents can be represented as \emph{incident patterns}. Incident patterns are a more abstract representation of specific incident instances. Thus they can be applied over different cyber-physical systems. They can also avoid disclosing potentially sensitive information about an organization's assets and resources (e.g., physical structure of a building or vulnerable devices). We also provided a representation of the system where an incident occurs. We provided two meta-models to represent incidents and cyber-physical systems, respectively. To obtain incident patterns, we proposed an automated technique to \emph{extract} an incident pattern from a specific incident instance. To understand how an incident pattern can manifest again in other systems, we proposed an automated technique to \emph{instantiate} incident patterns to different systems. We demonstrated the feasibility of our approach in the application domain of smart buildings. We evaluated correctness, scalability, and performance using two substantive scenarios inspired by real-world systems and incidents.

In future work we plan to apply our approach to support forensic readiness. Forensic readiness is the ability of a system to proactively identify and collect data that can be used in future investigations \cite{Rowlingson2004}. We intend to develop an automated technique that analyzes the output (i.e. traces) of the instantiation technique to identify system components and actions that can be relevant for future investigations. We also plan to apply our approach to various scenarios, for example, a scenario where an incident happens across different floors, and across smart buildings. Finally, we intend to facilitate the modeling of security incidents and cyber-physical systems by providing a graphical editor over our meta-models.

%
\IEEEpeerreviewmaketitle

\section*{Acknowledgment}

This work was partially supported by ERC Advanced Grant no.
291652 (ASAP) and Science Foundation Ireland grants
10/CE/I1855, 13/RC/2094 and 15/SIRG/3501.

\ifCLASSOPTIONcaptionsoff
  \newpage
\fi



%
%
%

\bibliographystyle{IEEEtran}
\bibliography{biblio}

%




\end{document}